\documentclass[prl,showpacs,twocolumn,amsmath]{revtex4}

\begin{document}

\title{Two-Setting Bell Inequalities for Many Qubits}
\author{Kai Chen$^{1}$}
\author{Sergio Albeverio$^{1}$}
\author{Shao-Ming Fei$^{2,3}$}
\affiliation{$^1$Institut f\"ur Angewandte Mathematik, Universit\"at Bonn, D-53115,
Germany\\
$^2$Department of Mathematics, Capital Normal University, Beijing 100037,
China\\
$^3$Max Planck Institute for Mathematics in the Sciences, D-04103 Leipzig,
Germany}

\begin{abstract}
We present a family of Bell inequalities involving \emph{only
two} measurement settings of each party for $N>2$ qubits. Our
inequalities include all the standard ones with fewer than $N$
qubits and thus gives a natural generalization. It is shown that
all the Greenberger-Horne-Zeilinger states violate the
inequalities maximally, with an amount that grows exponentially
as $2^{{(N-2)}/2}$. The inequalities are also violated by some
states that do satisfy \emph{all} the standard Bell
inequalities. Remarkably, our results yield in an efficient and
simple way an implementation of nonlocality tests of many
qubits favorably within reach of the well-established technology
of linear optics.
\end{abstract}


\maketitle

Quantum states can exhibit one of the most striking features of quantum
mechanics, producing remarkable correlations which are impossible within a
local realistic description based on the notion of Einstein, Podolsky, and Rosen
\cite{EPR1935}. Constraints on statistical correlations imposed by local
realism are termed Bell inequalities after the pioneering work of Bell
\cite{Bell1964}. Derivation of new and stronger Bell inequalities is one of
the most important and challenging subjects in quantum-information processing.
It is an essentially conceptual problem to find out to what extent a state can
rule out any possibly local realistic description, and thus certify its quantum
origin and true nonlocality. Violation of the inequalities
is very closely related to the extraordinary power of realizing certain tasks
in quantum information processing, outperforming its classical counterpart, such as
building quantum protocols to decrease communication complexity \cite{BZZ2002}
and making secure quantum communication \cite{Scarani0103}.

Since Bell's work there have appeared many important
generalizations, including the Clauser-Horne-Shimony-Holt (CHSH)
\cite{CHSH1969} and Mermin-Ardehali-Belinskii-Klyshko
(MABK) inequalities \cite{MABK}. A set of multipartite Bell
inequalities has been elegantly derived by Werner and Wolf and by
\.{Z}ukowski and Brukner (WWZB), by using two dichotomic
observables per site \cite{WW2001,ZB2002}. One usually refers
to such inequalities as ``standard" ones. Tailored for
high-dimensional systems, Bell inequalities are constructed in such a
way that each measurement can bear more than two outcomes
\cite{highdimension}. This further motivated successive
experimental verification of nonlocality
\cite{Zhaoetal2003,Waltheretal2005}. Moreover, the inequalities
can lead to a detailed classification of multipartite entanglement
\cite{YCPZ2003}, while the Greenberger-Horne-Zeilinger (GHZ)
states are shown to be the only
states that violate maximally the MABK inequalities
\cite{zqchen2004}. We refer to \cite{reviews} and references
therein for recent nice reviews.

However, Scarani and Gisin, and \.{Z}ukowski
\textit{et al.} find that there exists a family of pure $N>2$ qubit states which
\emph{escape} violation of the ``complete" set of Bell inequalities
\cite{SG2001,ZBLW2002} when considering a restricted setup
(two dichotomic measurements and full correlation functions among all the parties).
Note that such a restricted setup is sufficient to detect entanglement of any
pure \emph{bipartite} state and is known as Gisin's theorem
\cite{Gisintheorem}. A notable work
\cite{Popescu-RohrlichPLA92} shows that, for a fully entangled
$N$-partite pure state there exist some projective
measurements for $N-2$ parties such that one can still observe a
violation of the CHSH inequality for the remaining two particles. The
insight can be further proved to lead to a violation of
two-setting Bell inequalities and thus implies
Gisin's theorem for any number of qubits. However,
such a construction substantially relies on a localized
entanglement only between two parties with the help of all the
other parties. Furthermore, an amount with exponentially
increasing violation (which is a key character mainly coming from
true multipartite entanglement) is totally lost, as the maximal
violation is less than or equal to $\sqrt{2}$.
The family escaping violation is a subset of the
generalized GHZ states given by
$|\psi \rangle =\cos \alpha |0,...,0\rangle +\sin \alpha |1,...,1\rangle$
with $0\!\leq \!\alpha \!\leq \!\pi /4$.
It describes GHZ states \cite{GHZ1990} for $\alpha =\pi /4$.
For $\sin 2\alpha \leq 1/\sqrt{2^{N-1}}$ and $N$
\emph{odd}, these states are proved to satisfy all the standard
inequalities \cite{ZBLW2002}. This is rather surprising as they
are a generalization of the GHZ states which
maximally violate the MABK inequalities.

There has been also notable progress in deriving stronger Bell
inequalities by employing more measurement settings
\cite{WuZong2003,LPZB2004}, which can be violated by a larger
class of states, including the generalized GHZ states. Chen \textit{et al.}
recently obtained a Bell inequality for 3 qubits
involving two dichotomic observables per site, which can be seen
numerically to be violated by any pure entangled state
\cite{Chen-Wu-Kwek-Oh2004}. Can one find any Bell inequalities
satisfying the conditions that (i) they recover the standard Bell
inequalities as a special case; (ii) they provide an
exponentially increasing violation for GHZ states; (iii)
they essentially involve \emph{only two} measurement settings per
observer; iv) they yield violation for the generalized GHZ states
in the whole region of $\alpha$ for \emph{any number} of qubits?
This is highly desirable, as such Bell inequalities will lead to a
much easier and more efficient way to test nonlocality, and
contribute to the development of novel \emph{multiparty} quantum
protocols and cryptographic schemes by exploiting \emph{much less
entangled resources and experimental efforts.}

In this Rapid Communication, we present the first family of two-setting Bell
inequalities with all these advantages. We then show that it leads
to a natural generalization of the standard Bell inequalities. The
GHZ states are demonstrated to violate the inequalities maximally, by an
amount that grows exponentially as $2^{{(N-2)}/2}$. Finally,
we provide practical settings to test experimentally
the nonlocality of any generalized GHZ entangled states.

The scenario is as follows. We consider $N$ parties and allow each of them to
choose independently between two dichotomic observables $A_{j},A_{j}^{\prime
}$ for the $j$th observer, specified by some local parameters, each
measurement having two possible outcomes $-1$ and $1$. We define
\begin{widetext}
\begin{eqnarray}
\mathcal{B} &=&\mathcal{B}_{N-1}\otimes \frac{1}{2}(A_{N}+A_{N}^{\prime
})+\openone_{N-1}\otimes \frac{1}{2}(A_{N}-A_{N}^{\prime }),  \label{BI} \\
\mathcal{B}_{N-1} &=&\frac{1}{2^{N-1}}\sum_{s_{1},...,s_{N-1}=-1,1}
S(s_{1},...,s_{N-1})
\sum_{k_{1},...,k_{N-1}=1,2}
s_{1}^{k_{1}-1}...s_{N-1}^{k_{N-1}-1}\otimes _{j=1}^{N-1}O_{j}(k_{j}),
\label{bn-1}
\end{eqnarray}
\end{widetext}
where $\mathcal{B}_{N-1}$ is the quantum mechanical Bell operator of
WWZB inequalities \cite{WW2001,ZB2002}, and
$S(s_{1},...,s_{N-1})$ is an arbitrary function taking only
values $\pm 1$. Here $O_{j}(1)=A_{j}$ and $O_{j}(2)=A_{j}^{\prime }$ with $k_{j}=1,2$.
The notation $\openone_{N-1}$ represents an identity matrix of dimension $2^{N-1}$,
with the meaning of ``not measuring" the $N-1$ parties \cite{notm}.

From the classical view of local realism, the values of $A_{j},A_{j}^{\prime }$
are predetermined by a local hidden variable $\lambda $ before
measurement, and independent of any measurements, orientations or actions
performed on other parties at spacelike separation. The correlation among
all $N$ observations is then a statistical average over many runs of the
experiment
\begin{equation}
E_{\text{LHV}}({k_{1}},...,{k_{N}})={\int d\lambda \text{ } \rho (\lambda
)\prod_{j=1}^{N}O_{j}(k_{j},\lambda )},  \label{lcorr}
\end{equation}
where $\rho (\lambda )$ is a statistical distribution of $\lambda $
satisfying $\rho (\lambda )\geq 0$ and $\int d\lambda \text{ } \rho (\lambda )=1$.
Noting that local realism requires that $\left\vert \left\langle \mathcal{B}%
_{N-1}\right\rangle _{\text{LHV}}\right\vert \leq 1$ shown in \cite{ZB2002}, we obtain
\begin{equation}
\left\vert \left\langle \mathcal{B}\right\rangle _{\text{LHV}}\right\vert
\hspace{-0.3mm}=\hspace{-0.3mm}\frac{1}{2}\left\vert \left\langle \mathcal{B}_{N-1}(A_{N}+A_{N}^{\prime
})+(A_{N}-A_{N}^{\prime })\right\rangle _{\text{LHV}}\right\vert \leq 1.
\label{ourBI}
\end{equation}
In fact $A_{N}$ $=\pm 1$ and $A_{N}^{\prime }$ $=\pm 1$ for the observer $N$, and one
has either $|A_{N}+A_{N}^{\prime }|=2$ and $|A_{N}-A_{N}^{\prime }|=0$, or
vice versa. This implies that (\ref{ourBI}) holds. For a given function of
$S(s_{1},...,s_{N-1})$, one can generate the full set of members of a family
by simply permuting different locations, or the measurement orientations $A_{i}$
and $A_{i}^{\prime }$.

Let us now see the quantum-mechanical representation of the Bell
inequalities given by Eq.~(\ref{ourBI}) tailored for qubits. Since
any quantum observable $A_{i}$ that describes a measurement with
$\pm 1$ as possible outcomes can be represented by
$\vec{a}_{i}\cdot \vec{\sigma} \equiv \sigma _{a_{i}}$, with$\
\vec{a}_{i}$ a unit vector and $\vec{\sigma}$ the Pauli matrices
($A_{i}^{\prime }=\vec{a^{\prime }}_{i}\cdot \vec{\sigma} =\sigma
_{a_{i}^{\prime }}$, respectively), the Bell operator of
Eq.~(\ref{BI}) can be parametrized by all these $\sigma _{a}$. A
violation by the quantum state with density matrix $\rho $ reads
$\left\vert \left\langle \mathcal{B}\right\rangle \right\vert
=\left\vert \text{Tr}(\rho \mathcal{B}) \right\vert >1$. Moreover, every
unit vector $\vec{a}_{i}$ can be parametrized  completely by its
polar angle $\theta _{i}$ and azimuthal angle $\phi _{i}$ in the
Bloch sphere as $\vec{a}_{i}=(\sin \theta _{i}\cos \phi _{i},\sin
\theta _{i}\sin \phi _{i},\cos \theta _{i})$.

For the convenience of later use, we first derive an alternative form of
the MABK inequalities, different from the usual one through a recursive
definition \cite{MABK}. It is shown in \cite{ZB2002} that one can recover the
MABK inequalities by taking $S(s_{1},\ldots,s_{N})=\sqrt{2}\cos
[(s_{1}+\cdots+s_{N}-N+1)\pi /4]$ in Eq.~(\ref{bn-1}). Thus Eq.~(\ref{bn-1}) is
symmetric with respect to $s_{i}$, and one concludes
that the coefficients $c_{m}$ for
the correlation function $\otimes _{j=1}^{N}O_{j}(k_{j})$ will be
the same if the number of items for ${k_{i}=2}$ is fixed to be $m $. Without
loss of generality, supposing ${k_{i}=2}$ only for $i=1,\ldots,m$, we have
\begin{eqnarray}
c_{m}\hspace{-1mm} &=&2^{1/2-N}\hspace{-0.5mm}
\sum_{{s_{1},\cdots,s_{N}}\atop {=-1,1}}
\hspace{-0.5mm}\cos [(s_{1}+\cdots+s_{N}-N+1)\frac{\pi
}{4}] s_{1}\cdots s_{m}  \notag \\
&=&2^{1/2-N}{\mathop{\rm Re}\nolimits}\Big[e^{i(2m-N+1)\pi
/4}(e^{i\pi /4}+e^{-i\pi /4})^{N}\Big]  \notag \\
&=&2^{(1-N)/2}\cos [(2m-N+1)\pi /4],  \label{coef-MABK}
\end{eqnarray}
where we have used the fact that $s_{i}=\exp [i(1-s_{i})\pi /2]$ holds for
any $s_{i}=\pm 1$.

With the above notation and preparation, we can state several
main results.

\emph{Theorem 1. The generalized Bell inequalities Eq.~(\ref{ourBI}) include the standard
Bell inequalities as a special case.}

\textit{Proof.}
If one takes $A_{N}=A_{N}^{\prime }$, the inequalities reduce to $\left\vert \left\langle
\mathcal{B}_{N-1}\right\rangle _{\text{LHV} }\right\vert \leq 1$, which are precisely the
WWZB inequalities for $N-1$ parties. Furthermore, our inequalities inherit the
property that all the standard inequalities for fewer than $N-1$ parties will be
recovered, as it is valid for the standard inequalities \cite{WW2001}.
\hfill \rule{1ex}{1ex}

One may wonder whether all the GHZ states violate the
inequalities maximally, similarly in the case of MABK
inequalities. We provide the following answer.

\emph{Theorem 2. All the GHZ states violate the Bell inequality
Eq.~(\ref{ourBI}) maximally.}

\textit{Proof.} Squaring Eq.~(\ref{BI}) gives
\begin{eqnarray}
\mathcal{B}^{2} &=&\mathcal{B}_{N-1}^{2}\otimes \frac{1}{2}(1+\vec{a}
_{N}\cdot \vec{a^{\prime }}_{N})\openone_1  \notag \\
&&+\openone_{N-1}\otimes \frac{1}{2}(1-\vec{a}_{N}\cdot \vec{a^{\prime }}_{N})
\openone_1
\label{maxviolation}
\end{eqnarray}
Noting that $2^{N-2} \openone_{N-1} \geq \mathcal{B}_{N-1}^{2}$
as proved in \cite{WW2001}, one has $2^{N-2} \openone_{N} \geq
\mathcal{B}^{2}$. Here by $A \geq B$ we mean that $A-B$ is
semipositive definite. Thus a possible maximal violation is
$2^{(N-2)/2}$ due to the observation that the maximally possible
eigenvalue for $\mathcal{B}$ is $2^{(N-2)/2}$. This can indeed be
saturated by the GHZ states as seen from Eq.~(\ref{Bellviolation})
with $\alpha =\pi /4$, as will be shown in an example later. All
the other GHZ states up to a local unitary transformation will
lead to the same violation, since the case corresponds to a local
unitary transformation of local observables. \hfill \rule{1ex}{1ex}

As is well known the GHZ states are major resources for many
quantum-information tasks, while the generalized GHZ state are
crucial for distributed quantum computing \cite{YimLom2005}.
Moreover, in any real experiments for preparing the GHZ states, one
usually gets the generalized GHZ states due to unavoidable imperfections
in the devices. In the following we will find practical experimental settings
to detect this distinctive class of states. The correlation function is of the form
\begin{eqnarray}
\langle \otimes _{j=1}^{N}O_{j}(k_{j})\rangle &=&[\cos ^{2}\alpha +(-1)^{N}\sin
^{2}\alpha ]\prod_{i=1}^{N}\cos \theta _{i}  \notag \\
&&+\sin 2\alpha \prod_{i=1}^{N}\sin \theta _{i}\cos
\Big(\sum\limits_{j=1}^{N}\phi _{j}\Big),  \label{corrfunc}
\end{eqnarray}
as shown in \cite{WuZong2003}. Let us take $\theta _{i}^{1}=\theta
_{i}^{2}=\pi /2,$ $\phi _{i}^{1}=0,$ and $\phi _{i}^{2}=\pi /2$ for all
$i=1,\ldots ,N-1$, and set $\phi _{N}^{1}=\phi _{N}^{2}=\phi _{N}$ and
$\theta _{N}^{1}=\pi -\theta _{N}^{2}=\theta _{N}$. These special choices of
the angles correspond to measuring the first $N-1$ parties along $\sigma
_{x} $ (corresponding to $A_{i}$) and $\sigma _{y}$ basis ($A_{i}^{\prime }$).
Taking $\mathcal{B}_{N-1}$ as the MABK polynomial \cite{MABK} and using
Eqs.~(\ref{coef-MABK}) and (\ref{corrfunc}), one arrives at
\begin{eqnarray}
&&\left\langle \mathcal{B}_{N-1}\otimes \frac{1}{2}(A_{N}+A_{N}^{\prime
})\right\rangle =2^{(2-N)/2}\sin 2\alpha \text{ }\sin \theta _{N}  \notag \\
&&\hspace{3mm}\times \sum\limits_{m=0}^{N-1}{\binom{N-1}{m}}\cos \Big((2m-N+2)
\frac{\pi }{4}\Big)\cos \Big(\frac{m\pi }{2}+\phi _{N}\Big)  \notag \\
&&=2^{(N-2)/2}\sin 2\alpha \text{ } \sin \theta _{N}\sin \Big(\frac{N\pi }{4}+\phi _{N}\Big)
\notag \\
&&=2^{(N-2)/2}\sin 2\alpha \text{ } \sin \theta _{N},  \label{corr1}
\end{eqnarray}
which can be easily derived by using exponential representations of
trigonometric functions, and noting that the sum corresponds to the binomial
expansion of a certain function. Here $m$ is the number of the parties that
are measured along the $\sigma _{y}$\ basis. In the last formula of
Eq.~(\ref{corr1}), we have further set $\phi _{N}=(2-N)\pi /4$.

It is easy to see that $\left\langle \openone_{N-1}\otimes (A_{N}-A_{N}^{\prime
})/2\right\rangle =\cos 2\alpha \text{ } \cos \theta _{N}$, one thus has
\begin{equation}
\left\langle \mathcal{B}\right\rangle =2^{(N-2)/2}\sin 2\alpha \text{ } \sin \theta
_{N}+\cos 2\alpha \text{ } \cos \theta _{N}.  \label{violationgGHZ}
\end{equation}
Since $\max (x\sin \theta +y\cos \theta )=\sqrt{x^{2}+y^{2}}$, we get
\begin{eqnarray}
\left\langle \mathcal{B}\right\rangle &=&(2^{N-2}\sin ^{2}2\alpha +\cos
^{2}2\alpha )^{1/2}>1  \label{Bellviolation} \\
&&\hspace{3mm}\text{for }\alpha \neq k\pi /2\text{ \ \ \ }(k\in \text{integer}
,N\geq 3),  \notag
\end{eqnarray}
where we have taken $\theta _{N}=\tan ^{-1}\big(2^{(N-2)/2}\tan 2\alpha \big)$ if
$0\leq \alpha \leq \pi /4$, and $\theta _{N}=\tan ^{-1}\big(2^{(N-2)/2}\tan
2\alpha \big)+\pi $ if $\pi /4\leq \alpha \leq \pi /2$ in
Eq.~(\ref{violationgGHZ}). Therefore, we see that the whole class of generalized GHZ
states can violate the Bell inequalities Eq.~(\ref{ourBI}) except for the
product states ($\alpha =0$ or $\pi /2$).

By suitable choices of the two observables in each site, one thus can
reveal hidden nonlocality for any generalized GHZ state
in a very subtle way through our inequalities. Note that the result
leads to \emph{the same violation factor} as the one obtained by
the \emph{many settings} approach \cite{LPZB2004}, where however the
required experimental effort is \emph{exponentially larger} than ours.
In addition, for a given $\alpha$ the violation will increase exponentially, with the
maximal violation achieved by GHZ states.

Let us highlight the significance of our inequalities. First,
the implementations involve only two measurement settings per
site, and should be immediately feasible due to rapidly developing
technology for generation and manipulation of multipartite
entangled states in linear optical, atomic, or trapped ion systems
\cite{Zhaoetal2003,Leibfried-Hafner2005}. Second, the standard
Bell inequalities are recovered as a special case of our
inequalities as shown in Theorem 1. Third, for $N$ even our
inequalities demand asymptotically only half of the experimental
efforts. In such a case the MABK inequalities are combinations of
all the correlation functions with $2^{N}$ terms \cite{MABK}. Our
inequalities require only $2^{N-1}+2$ terms, as seen from Eq.~(\ref{BI})
($\mathcal{B}_{N-1}$ is a combination of $2^{N-2}$ correlation
functions in this case). Fourth, they fill the well-known gap for
the states that the standard Bell inequalities fail to detect,
and keep the exponentially increasing violation in the mean time.

From Eq.~(\ref{ourBI}), one can see that our inequalities not only
include the full correlations, but also account for
fewer than $N$ particle contributions. This novel construction
goes beyond the restricted set classified in \cite{WW2001,ZB2002}, and
exhibits superior power by admitting a wider class of LHV description to
include all possible correlations, as was done before for three qubits in
\cite{Chen-Wu-Kwek-Oh2004}, and for four qubit cluster states in
\cite{Scaranietal2005} with a linear optical demonstration in
\cite{Waltheretal2005}.

We remark that our inequalities apply as well to
arbitrary dimensional multipartite systems. Moreover, they can be violated by
a $|W\rangle $ state of the form $(1/\sqrt{N})(|100...0\rangle
+|010...0\rangle +\cdots+|000...1\rangle )$, and by cluster states that
are effective resource for one-way universal quantum computation
\cite{Raussendorf-Briegel2001}. For example, taking $\mathcal{B}_{N-1}$ as
the MABK polynomial in Eq.~(\ref{BI}), the $|W\rangle $ state can
be violated with maximal violation factors of $1.202,1.316,1.382$
for $N=3,4,5$, respectively, while for
the cluster states $|\psi _{3}\rangle =(1/\sqrt{%
2})(|000\rangle +|111\rangle )$ (GHZ state), $|\psi _{4}\rangle
=(1/2)(|0000\rangle +|0011\rangle +|1100\rangle -|1111\rangle )$
\cite{Scaranietal2005,Waltheretal2005} a factor of $\sqrt{2}$ for
both. Considering a practical noise admixture to the $N$-particle GHZ
state $|\text{GHZ}\rangle$ of the form $\rho = (1-V) \rho_{\text{noise}} +
V |\text{GHZ} \rangle \langle \text{GHZ}|$, with $\rho_{\text{noise}} = \openone
/ 2^N$, one has a threshold visibility
of $V_{\text{thr}}=2^{(2-N)/2}$ above which a local realism is
impossible. This suggests that our inequalities are rather
efficient, as it is only a slightly bigger threshold visibility
than that required for the MABK inequalities with
$V_{\text{thr}}=2^{(1-N)/2}$. In addition, they share the same
behavior as $V_{\text{thr}}\sim 2^{-N/2}$ that is exponentially
decreasing to 0 in the asymptotics $N\rightarrow \infty$.
For $N\geq 4$, they are also significantly better than the one
derived from \cite{Popescu-RohrlichPLA92} which requires the very
strict condition $V_{\text{thr}}\geq 0.7071$ for any $N$.

Summarizing our results, we have proposed a novel family of Bell
inequalities for many qubits. They are entirely compatible with
the simplicity requirements of current linear optical experiments
for nonlocality tests, i.e., involving only two measurement
settings per location. The inequalities recover the standard
Bell's inequalities as a special case and can be maximally
violated by GHZ states. In addition, practical experimental
settings are derived for revealing violation of local realism
for some class of states which the standard Bell's inequalities fail
to detect. This permits to reduce significantly experimental
efforts comparing with those which utilize many settings, and at
the same time, can be achieved without compromise of an exponentially
increasing amount of violation. Complementary to
the standard inequalities and a number of existing results, our
inequalities offer another prospective tool for much \emph{stronger}
nonlocality tests and a \emph{more economic} way of performing
experiments.

We thank Jian-Wei Pan, Zeng-Bing Chen, Chunfeng Wu and Jing-Ling Chen for
valuable discussions and communications. K.C. gratefully acknowledges support
from the Alexander von Humboldt Foundation. The support of this work by the
Deutsche Forschungsgemeinschaft SFB611 and the German (DFG)-Chinese (NSFC) Exchange
Programme No. 446CHV113/231 is also gratefully acknowledged, as well as the partial
support by NKBRPC (Grant No. 2004CB318000).

\end{document}